\documentclass[acmlarge,screen,nonacm]{acmart}
\setcopyright{none}
\renewcommand\footnotetextcopyrightpermission[1]{}
\acmYear{2026}
\AtBeginDocument{}
\begin{document}
\title[Iterative Design and Evaluation of a VR Nursing Training System]{Iterative Design and Evaluation of a VR Nursing Training System: Insights from Clinicians and Student Learners}
\author{Danhua Zhang}
\email{zhan5954@umn.edu}
\affiliation{\department{Department of Computer Science and Engineering}\institution{University of Minnesota Twin Cities}\country{USA}}

\author{Cynthia Bradley}
\email{cbradley@umn.edu}
\affiliation{\department{School of Nursing}\institution{University of Minnesota Twin Cities}\country{USA}}

\author{Christopher Kissell}
\email{kiss0027@alumni.umn.edu}
\affiliation{\department{School of Nursing}\institution{University of Minnesota Twin Cities}\country{USA}}

\author{Courtney Hutton Pospick}
\email{hutto070@umn.edu}
\affiliation{\department{Department of Computer Science and Engineering}\institution{University of Minnesota Twin Cities}\country{USA}}

\author{Seraphina Yong}
\email{seraphinacyong@gmail.com}
\affiliation{\department{Department of Computer Science and Engineering}\institution{University of Minnesota Twin Cities}\country{USA}}

\author{Qiao Jin}
\email{qjin4@ncsu.edu}
\affiliation{\department{Department of Computer Science}\institution{NC State University}\country{USA}}

\author{Tongyu Nie}
\email{nie00035@umn.edu}
\affiliation{\department{Department of Computer Science and Engineering}\institution{University of Minnesota Twin Cities}\country{USA}}

\author{Evan Suma Rosenberg}
\email{suma@umn.edu}
\affiliation{\department{Department of Computer Science and Engineering}\institution{University of Minnesota Twin Cities}\country{USA}}

\renewcommand{\shortauthors}{Zhang et al.}

\authorsaddresses{Authors' Contact Information: Danhua Zhang (corresponding author), Department of Computer Science and Engineering, University of Minnesota Twin Cities, USA, e-mail: \href{mailto:zhan5954@umn.edu}{zhan5954@umn.edu}; Cynthia Bradley, School of Nursing, University of Minnesota Twin Cities, USA, e-mail: \href{mailto:cbradley@umn.edu}{cbradley@umn.edu}; Christopher Kissell, School of Nursing, University of Minnesota Twin Cities, USA, e-mail: \href{mailto:kiss0027@alumni.umn.edu}{kiss0027@alumni.umn.edu}; Courtney Hutton Pospick, Department of Computer Science and Engineering, University of Minnesota Twin Cities, USA, e-mail: \href{mailto:hutto070@umn.edu}{hutto070@umn.edu}; Seraphina Yong, Department of Computer Science and Engineering, University of Minnesota Twin Cities, USA, e-mail: \href{mailto:seraphinacyong@gmail.com}{seraphinacyong@gmail.com}; Qiao Jin, Department of Computer Science, NC State University, USA, e-mail: \href{mailto:qjin4@ncsu.edu}{qjin4@ncsu.edu}; Tongyu Nie, Department of Computer Science and Engineering, University of Minnesota Twin Cities, USA, e-mail: \href{mailto:nie00035@umn.edu}{nie00035@umn.edu}; Evan Suma Rosenberg, Department of Computer Science and Engineering, University of Minnesota Twin Cities, USA, e-mail: \href{mailto:suma@umn.edu}{suma@umn.edu}.}

\subtitle{Preprint. This manuscript is under review.}
\begin{teaserfigure}
\includegraphics[width=\textwidth]{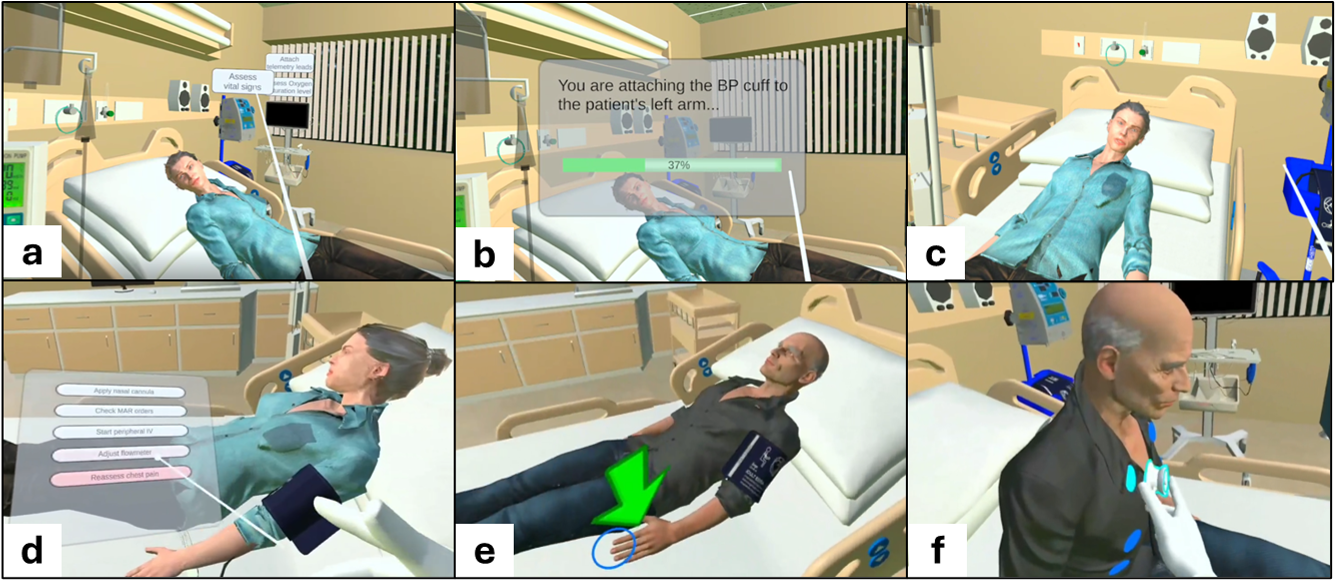}
\caption{Screenshots from the chest pain and oxygen therapy scenarios used in the two studies. In the first study, (a) the treatment buttons are located near the corresponding medical devices; (b) the progress bar appears in front of the view upon button selection; (c) verbal communication with the patient in the chest pain scenario. In the second study, (d) the treatment buttons are located on the same panel; (e) users need to grab and place the medical instruments in the correct location upon button selection as indicated by the visual cues; (f) manual interaction (checking lung sounds) with the patient in the oxygen therapy scenario.}
\Description{Six screenshots show the chest pain and oxygen therapy scenarios, including treatment buttons, a progress bar, patient dialogue, an action panel, placement cues, and lung auscultation.}
\label{fig:teaser}
\end{teaserfigure}
\begin{abstract}
This paper presents the iterative design and evaluation of a virtual reality (VR) nursing training system developed in collaboration with nursing education specialists to enhance procedural skill acquisition, clinical decision-making, and critical thinking among pre-licensure nursing students. Two clinical scenarios—chest pain and oxygen therapy—were created to simulate acute care settings, with design refinements informed by an expert evaluation study (n=8) and further assessed through an effectiveness study involving nursing students (n=17). Results demonstrate high user acceptance, perceived usefulness, and significant post-simulation increases in learner confidence. The VR system was found to be an accessible supplemental training tool that can foster self-directed learning. Finally, we present design recommendations for developing VR simulations for nursing education, emphasizing collaboration with domain experts, accommodating diverse learning goals, and simulating complex clinical workflows.
\end{abstract}
\maketitle

\section{Introduction}

Providing novice nurses with adequate clinical experience remains a challenge for nurse educators due to the scarcity of available clinical sites, which consequently curtails hands-on practice time. 
In conventional training settings, students typically practice in a clinical skills lab or operating room alongside peers, often with limited direct supervision. 
Although faculty assistants may provide occasional support, the absence of clear, step-by-step guidance can leave students uncertain about where to begin—particularly due to the overwhelming array of medical devices and equipment in the environment \cite{mansoorian2019exploring}.
This has led to the widespread adoption of simulation-based learning, formally defined as an educational intervention which creates an environment that is conducive to experiential learning \cite{norman2012systematic}. 

Nursing simulations can be broadly categorized into five main types based on the interface: standardized patient actors \cite{barrows1993overview}, mannequins ranging from low- to high-fidelity models \cite{kilmon2010immersive}, desktop-based virtual patient interactions \cite{choi2016emerging, KLEINSMITH2015151}, immersive virtual reality (VR) environments \cite{foronda2017virtually, KARDONGEDGREN201928}, and augmented reality (AR) applications for mobile devices and head-mounted displays (HMDs) \cite{wuller2019scoping}. 
While methods like standardized patients and mannequin-based simulations provide opportunities for hands-on skill practice, they have notable limitations. 
Patient actors must be trained extensively to portray scenarios consistently, introducing an undesirable degree of variability for training and testing medical knowledge and procedural skills. 
Mannequins allow psychomotor skill training, but most lack the dynamic responses and verbal interactions of real patients. 
Those that provide these interactions are often cost-prohibitive \cite{kilmon2010immersive}. 
Virtual simulations mitigate these constraints by offering standardized, reproducible scenarios with dynamic responses calibrated precisely and consistently for learning objectives. 
Immersive VR uniquely enables nursing students to engage with realistic simulations replicating the visual, auditory, and spatial dimensions of care environments. 

Virtual simulations are more cost-effective than mannequin-based approaches, with the cost-utility ratio per student estimated at \$1.08 for virtual simulations compared to \$3.62 for mannequin-based exercises \cite{haerling_cost-utility_2018}. 
Some estimates for high-fidelity mannequins place the cost-utility ratio as high as \$6.28 per student \cite{lapkin2011cost}. 
Unlike desktop-based modules, immersive VR allows higher degrees of interactivity and immersion \cite{kilmon2010immersive} while maintaining a low investment cost. 

This research focused on applying VR as a supplemental training tool for developing nursing students' integrated medical knowledge and clinical procedural skills. 
We are aiming to explore the following research questions:
\begin{itemize}
    \item \textbf{(RQ1)} How does iterative design collaborating with nursing specialists shape the development of a VR training system?
    \item \textbf{(RQ2)} How do nursing students perceive the usability and effectiveness of the designed VR training system?
    \item \textbf{(RQ3)} What key factors should designers prioritize when designing VR training systems in collaboration with nursing specialists?
\end{itemize}

We introduce related previous works in Section \ref{related-work}, describe the two iterative phases of system design and procedure in Section \ref{sec:co-design-process}. First, we designed a scenario where a patient has chest pain to establish the framework, followed by an expert evaluation study engaging nursing experts ($n=8$) to review the system and provide feedback through hands-on sessions and interviews, enabling iterative refinements. 
Next, we extended the system to a second scenario where a patient needs oxygen therapy to demonstrate the adaptability of the system followed by the second study ($n=17$) with nursing students indicating the revised VR simulations achieved high efficacy and user acceptance, enhancing learners' self-confidence. 
Both studies are mixed-methods and have been approved by the Institutional Review Board (IRB) of the local institution. Synthesizing insights from both studies, we provide a set of design recommendations tailored to developing VR procedural training for healthcare education contexts in Section \ref{sec:discussion}. 
For clarity, we distinguish between (1) ``nursing education specialists", the two collaborators engaged across the design process, and (2) ``nursing experts," participants who conducted the preliminary evaluations in Study 1.
The major contributions of this work include:
\begin{itemize}
    \item We describe a prototype VR nursing training system that was iteratively designed with nursing education specialists: initial design and refinement, which can be extended to similar clinical scenarios, supporting procedural skills, critical thinking, and decision making.
    \item We present empirical findings from two user studies conducted with nursing experts and pre-licensure nursing students, respectively.
    \item We summarized lessons learned and design considerations that were identified during the design process and from the feedback of the participants. 
\end{itemize}

\section{Related Works} \label{related-work}
Virtual reality (VR) has become an increasingly prominent tool in healthcare training, offering immersive, reproducible, and cost-effective simulations that overcome the limitations of traditional methods such as standardized patients and mannequins. By enabling realistic visual, auditory, and spatial engagement, VR has shown promise across diverse healthcare domains, from surgical training to clinical decision-making.

\subsection{VR Simulation in Healthcare}
In the broader healthcare context, VR has been widely employed in medical and surgical training. Early clinical studies demonstrated that VR-based interventions could improve operating room performance and accelerate skill acquisition compared to conventional training methods \cite{mao2021immersive}, such as robot surgery \cite{bric2016current}, laparoscopic surgery \cite{alaker2016virtual}, plastic surgery \cite{bielsa2021virtual}. Beyond surgery, VR-based virtual patients and team simulations have been applied in fields such as emergency medicine, disaster preparedness, and interprofessional collaboration, where they provide safe, repeatable environments for high-stakes training \cite{kononowicz2015virtual, shujuan2022use}. Much of the existing literature on virtual simulation-based training for healthcare practitioners has concentrated on improving communication skills, often employing conversational virtual patients to facilitate learning \cite{fernandez2025virtual}. These studies have predominantly targeted medical students or physicians, with comparatively less attention given to nursing education. Among the research that does focus on nurses, most training systems have been implemented using desktop-based virtual environments, using nursing scenarios such as pediatric interviews \cite{dukes2013sidnie} and triage assessment \cite{stuart2021building}.  Social virtual humans have also been investigated for conflict resolution during team training with nurses and surgeons, leveraging large-format displays installed in an operating room \cite{robb2015comparison}. More recently, researchers have explored the use of augmented reality training for triage assessment, signaling a growing interest in expanding the modalities of virtual simulation for nursing education \cite{anderson2021augmented}.

The system presented in this paper leverages immersive virtual reality to support the training of procedural skills in nursing education.
\color{black}
Prior research has primarily investigated the feasibility, acceptability, and usability of VR simulations in nursing education, either through custom-developed interventions or commercial platforms like vSim \cite{GU2017194} and Ubisim \cite{4357969, DERRICO202130}. 
Studies such as Schuelke et al.'s \cite{schuelke2022implementing} examination of a COVID-19 patient scenario, Saab et al.'s \cite{saab2022nursing} exploration of testicular disease awareness training, and Botha et al.'s \cite{BOTHA202176} analysis of airway obstruction management, and Butt et al.'s \cite{BUTT201825} evaluation of a game-based catheterization system consistently reported positive student attitudes toward VR adoption. 
These works demonstrated VR's potential for engaging learners in targeted clinical scenarios, ranging from airway management to procedural skills training. 

A separate body of research has evaluated the effectiveness of VR simulations through measurable metrics including clinical performance, self-confidence, self-efficacy, nervousness, and anxiety reduction \cite{chen2020effectiveness, foronda2020virtual, SHOREY2021104662, jallad2021effectiveness, PLOTZKY2021104868, choi2022effectiveness, TAMILSELVAN2023105711}. 
For instance, Sen et al.'s \cite{SEN2022199} analysis of surgical aseptic training found VR groups outperformed traditional methods while reporting lower anxiety levels.
Liao et al.'s \cite{shujuan2022use} disaster response simulation and Adhikari et al.'s \cite{ADHIKARI2021104944} sepsis game intervention demonstrated significant improvements in students' clinical preparedness, confidence, and skill acquisition, with the latter specifically showing enhanced self-efficacy and reduced anxiety through its game-based approach.

However, technical implementation details remain scarce in nursing VR literature, as most studies either employ commercial simulation solutions without disclosing system architectures, or develop symptom-specific scenarios that lack generalizability. This opacity stems from the field's predominant publication in medical venues emphasizing learning outcomes over technical design considerations.
Although some recently published work on VR-based training for nursing education has emphasized technical implementation, multi-stakeholder design perspectives for clinical training applications were not addressed \cite{obeid2025nursingxr}. 
In contrast, our work highlights the design perspective of the simulation, demonstrates how to extend the system to another scenario, and provide design guidance for building similar applications.
\subsection{Procedural Skills Training}
VR simulations have been used to train a wide range of healthcare competencies, including clinical skills, procedural and psychomotor skills, communication skills,
reasoning and judgment skills, and technical skills \cite{choi2022effectiveness}. Our VR simulation was specifically designed to facilitate nursing students' practice of procedural skills while integrating decision-making and critical-thinking components--key competencies for clinical readiness \cite{criticalthinking}.

Procedural skills training emphasizes mastery of step-by-step interventions, where adherence to correct sequences takes precedence over tactile or motoric fidelity. This focus reduces the complexity and cost of developing simulations compared to psychomotor skill trainers, offering inherent advantages such as repeatable practice and instructor-independent learning. However, while motoric capabilities and haptic feedback remain critical for psychomotor skill acquisition (e.g., catheter insertion or injection techniques), few existing simulations incorporate these features due to the high development costs and specialized hardware requirements (e.g., force-feedback devices). This gap highlights a tension in VR-based training design: balancing accessibility for procedural learning with the fidelity needed for nuanced psychomotor tasks. \cite{PLOTZKY2021104868}. 

Several previous research studies evaluated different VR simulations that were designed for procedural skill training. 
Bracq et al. \cite{BRACQ2019153} conducted a study aimed to assess the acceptability and usability of the VR simulator for procedural skill training among scrub nurses. 
Choi \cite{8797741} developed a virtual simulation of wound care training and ran an initial study with a registered nurse and a nursing student who had knowledge of change wound dressing to obtain feedback on the simulation.
Chang and Lai \cite{CHANG2021104670} conducted a focus group interview among students after students tried the VR simulation of nasogastric tube care skills.
Taçgın \cite{taccgin2020perceived} conducted a study to investigate the features of a virtual learning environment in VR and evaluate its effectiveness.
However, the above-mentioned VR systems were either narrowly focused on specific scenarios or failed to adopt a comprehensive design perspective that considers both nursing educators' pedagogical needs and students' learning experiences. 
Distinct from these approaches, our work advances the field by integrating technical system development with evidence-based design considerations, achieved through iterative collaboration with nursing education specialists.
By centering our development on this dual-stakeholder design philosophy, we bridge a critical gap between educational theory and VR system design in nursing training.

\section{Design Iteration of Framework: Phase 1} \label{sec:co-design-process}
\subsection{Requirements Gathering}
Our overarching goal was to design a system applicable to a wide range of nursing scenarios, enabling virtual patient interactions for procedural skill practice, clinical decision-making, and critical thinking development.
To achieve this, we formed an interdisciplinary team comprising two VR researchers (the first and last authors) and two nursing education specialists (the second and third authors) from [anonymized] University. 
The nursing collaborators brought dual expertise as both educators and researchers in reflective debriefing techniques, ensuring alignment between technical development and pedagogical best practices.
In this initial phase, our goal was to develop a proof-of-concept version of the system, prioritizing the key procedural elements necessary to ensure alignment with real-world nursing workflows and educational objectives. 
This foundational prototype was designed to validate core functionality while providing a flexible framework for future scenario expansion and refinement.

The nursing collaborators initially focused on a chest pain scenario because this is one of the most commonly encountered patient conditions in acute care settings.
Also, this scenario requires all the desired skills for nurses, including procedural skills, critical thinking, and decision making. 
In this scenario, trainees were tasked to relieve the chest pain of the virtual patient by interacting with and examining the patient in the correct order and administering the correct medications.
The nursing collaborators researched related medical literatures and teaching resources to make sure all the tests were accurate; for example, the speed of the oxygen flow speed to be adjusted, heart rate, respiratory rate, blood pressure etc. 
Based on that, they developed the medical materials, which included the virtual patient’s background, medical history, treatment objectives, and structured tasks. 
These tasks were decomposed into fundamental procedural steps to facilitate systematic learning. 

The scenario’s logic was designed through iterative collaboration between the nursing specialists and VR researchers, with a detailed flowchart mapping all possible clinical decision pathways (see supplemental materials). 
The scenario incorporates guided decision points, enabling learners to navigate tasks more intuitively while maintaining engagement.
Nursing collaborators meticulously evaluated each decision pathway, designing context-aware feedback for both correct and incorrect treatments. 

Beyond merely displaying symptoms in response to errors, the nursing collaborators suggested incorporating guided hints to steer learners toward corrective actions, thereby reinforcing experiential learning. 
Additionally, the team defined optimal procedural outcomes to ensure alignment with clinical best practices for symptom management. 
Throughout development, the system underwent rigorous evaluation by both VR researchers and nursing specialists, with iterative testing to refine interactions, validate medical accuracy, and optimize the learning experience.
At the end of this phase, the system consists of the following essential elements:
\begin{enumerate}
    \item A list of optional treatments derived from the task in a complex scenario, each corresponding to a sequence of interactions involving the patient, medical devices, or instruments. 
    \item A virtual patient who responds verbally to user-administered treatments. Designed for student self-practice, the virtual character also provides corrective hints when errors occur.
    \item A 3D emergency room (ER) with medical devices necessary in a specific scenario. In the chest pain scenario, the virtual room was equipped with a computer to give the medicine order, a sink, a peripheral IV device, a bed, a nasal cannula, an oxygen flow meter, a vital sign machine with a blood pressure cuff, and a 12-lead ECG device with an oximeter.
\end{enumerate}

\subsection{Prototype Design}
We proposed a prototype system based on the goal of training nursing students in procedural skills in VR.
\subsubsection{Apparatus}
An Oculus Quest headset and controllers were used for this VR simulation because mobile VR systems are portable and have a comparatively low price compared to PC-tethered systems, which would enable nursing students to perform self-training outside of class without the need for a personal computer equipped with a high-end graphics card. 
The VR simulation was made in Unity 2020.3, the patient avatar was developed in Reallusion Character Creator 3, and the facial animations were created in Reallusion iClone 7. 
The audio clips of the patient's verbal feedback were generated using Microsoft Azure.

\subsubsection{VR Simulation}
The VR simulation begins by displaying the scenario description and virtual patient's medical history in text form. 
Users are spawned at the door of the virtual ER and asked to release the chest pain of the
virtual patient based on the instructions and the patient’s feedback.
After reading, the user selects their dominant hand and initiates the simulation. 
During the simulation, the virtual patient would lie in the bed, engage in eye contact, and respond verbally to the user’s actions using viseme animations.

Users can navigate in the virtual room by head-directed steering walking and physical turning or snap-turning using either joystick. 
At each step, available treatments are visualized
as buttons beside the relevant objects and selected using the laser pointer by pressing the trigger button (see \ref{fig:teaser} a).
Users were asked to select the most appropriate treatment to be performed at each step, and must select treatments in the correct sequence to address the patient's symptoms, with available options displayed as buttons adjacent to the corresponding medical devices.
The dialogue system architecture employs a finite state machine model.
Each state contains a linked list of actionable user options corresponding to evaluation checkpoints, storing both selectable actions and their associated animations and next state IDs.
We implemented the state machine using Unity Coroutines for their asynchronous execution capabilities which are suitable for pre-built animations.

When each treatment was selected, the system would provide three responses: 
\begin{enumerate}
    \item The virtual patient would provide verbal feedback on their feelings, or hints to users of what they should do in the next step.  For example, if the user started measuring the patient without identifying their name, the patient would say: \textit{``Do you even know who I am? Don't you want to know my name?"} If the user only measured heart rate and blood pressure, the patient would say: \textit{``My chest pain isn't going away. Can’t you give me some medicine?” }
    
    \item A progress bar would appear to indicate the action the user was supposed to take. The purpose of this design is to make users focus on medical knowledge related to the sequential order of procedural skill and avoid distracting the users.

    \item For actions involving using the medical devices such as putting the blood pressure (BP) cuff on the virtual patient, the physiological data would appear or change accordingly on the device screen.
\end{enumerate}
The simulation concludes when the user completes all required treatments, as indicated by the virtual patient's verbal confirmation of symptom relief. For further details, please see the supplemental materials.

\subsection{Study 1: Expert Evaluation}
An expert evaluation study was conducted to assess the initial system version, identifying improvement opportunities and informing the system design for the next phase. 

\subsubsection{Participants}
Eight nursing experts were recruited from the School of Nursing of the local institution. 
75\% of them had a teaching role. 
The age of the participants ranged from 26 to 39 years ($Mean$ = 29.88, $SD$ = 5.59). 
All participants were female, right-handed, had either normal vision or vision corrected to normal by contact lens, and had an active license to practice nursing at the level of registered nurse or higher. 
Five participants did not play 3D video games, two had 10+ years of experience playing 3D video games, and one had less than a year of experience with 3D video games. 

\subsubsection{Procedure \& Analysis}
The participants were invited to a private study room and shown the information sheet. 
Before the simulation, they were given a brief introduction of the VR system and the task they need to complete in the simulation with screenshots of the scenario. 
Then they were asked to complete the Simulator Sickness Questionnaire (SSQ), put on the VR headset, adjust the head strap and interpupillary distance (IPD) and hold the controller in their dominant hand. 
After being familiar with the interface, they read the scenario description and the patient's medical background in VR and completed the simulation, followed by the post-study questionnaires and a semi-structured interview. 
The last step was to complete the demographics questionnaire to avoid any potential bias. 
The time for each session was about an hour and the compensation for each participant was a \$20 Amazon.com gift card. 

We employed the SSQ \cite{kennedy1993simulator} to assess potential adverse experiences, measuring participants' subjective cybersickness levels pre- and post-simulation.
To specifically identify motion sick participants, we also added the following yes/no question at the beginning of the questionnaire, ``Are you motion sick now?”. 

The User Acceptance Questionnaire (UAQ) \cite{Rasimah_Ahmad_Zaman_2011} was used aiming to measure user acceptance of using this simulation in
nursing education. It consists of five metrics: Personal Innovativeness (PI) \cite{agarwal1998conceptual}, Perceived Enjoyment (PE) \cite{Venkatesh2003}, Perceived Ease of Use (PEOU), Perceived Usefulness (PU) and Intention to Use (ITU) \cite{Davis1989}. 
PE utilized 6 items while other metrics used 4 items each. 
Since our target users are nursing students, we also modified the items applied to nursing students and added them to the UAQ. 
For example, an item ``\textit{using the simulation app would improve \textbf{my} performance in dealing with the chest pain scenario}" was modified to ``\textit{using the simulation app would improve \textbf{nursing students}' performance in dealing with the chest pain scenario}". 
Non-applicable items such as the PI metric were not modified.
We recorded simulation duration to estimate time requirements and conducted semi-structured interviews \cite{gill2008methods} to collect comprehensive feedback.

\subsection{Study 1: Results} \label{sec:result-study1}
\paragraph{Cybersickness}
No participants were self-identified motion sick before the study, and only one participant was self-identified to be sick after completing the simulation. 
The mean of the delta overall SSQ scores is 3.74 ($SD$ = 9.59) out of a maximum possible score of 235.62. 
The Wilcoxon Signed-Rank Test was used because the normality assumption was violated as indicated by the Shapiro-Wilk normality test. 
No significant difference was found between the pre-study and the post-study SSQ scores ($p$ = .20).
The results show that the system caused only mild adverse effects such as occasional blurry vision, likely attributable to the required mask and eye mask use during COVID-19 precautions.

\begin{figure*}[t]
    \centering
    \begin{minipage}{0.48\textwidth}
        \centering
        \includegraphics[width=\textwidth]{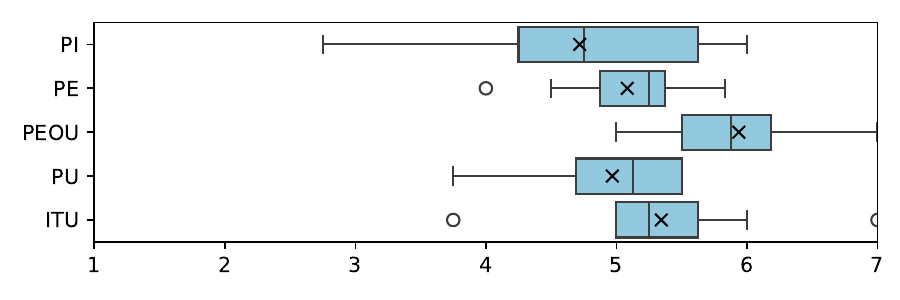}
        \label{fig:uaq-subscale-expert}
    \end{minipage}
    \hfill
    \begin{minipage}{0.48\textwidth}
        \centering
        \includegraphics[width=\textwidth]{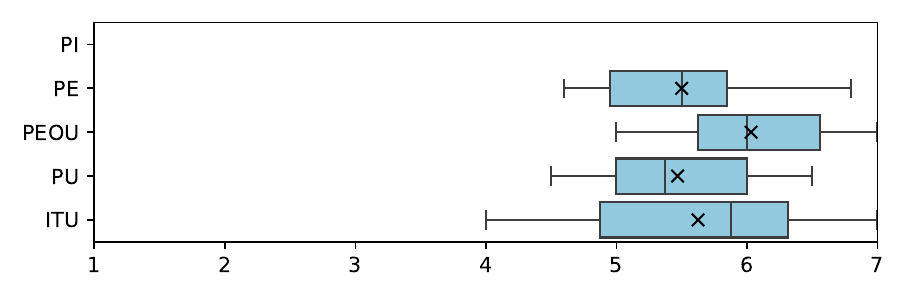}
        \label{fig:uaq-subscale-student}
    \end{minipage}
    \caption{Box plots of UAQ-subscale measures from Study 1: Personal Innovativeness (PI), Perceived Enjoyment (PE), Perceived Ease of Use (PEOU), Perceived Usefulness (PU) and Intention to Use (ITU), displaying the mean (cross marks), median, and IQR. The plots show experts' \textit{self-reported ratings} (left) and their perceptions of \textit{nursing students' ratings} (right).}
    \Description{Two sets of horizontal box plots summarize user acceptance ratings in Study 1 for experts and their perceptions of student ratings.}
\label{fig:UAQ-subscale-study1}
\end{figure*}

\paragraph{Subjective Questionnaire Responses}
The range of the scale in the UAQ is from 1 to 7 (strongly disagree = 1, strongly agree = 7). 
There were no PI items corresponding to nursing students. 
The data of the item "\textit{in general, I am hesitant to try out new information technologies} was reversed before computation. The overall mean of all PI items was 4.72 ($SD$ = 1.15). 
The item "\textit{I was unhappy when the simulation was over}" was not applied to nursing students. 
The overall mean of all PE, PEOU, PU, ITU items was 5.08 ($SD$ = 0.58), 5.94 ($SD$ = 0.66), 4.97 ($SD$ = 0.62), 5.34 ($SD$ = 0.93) respectively, and the overall mean of all of the items corresponding to nursing students was 5.50 ($SD$ = 0.73), 6.03 ($SD$ = 0.70), 5.47 ($SD$ = 0.69), 5.63 ($SD$ = 1.04) respectively. 
The results are illustrated in Fig. \ref{fig:UAQ-subscale-study1}.  
It shows that the participants generally held positive attitude towards using this system, and students.
In addition, the results of the UAQ showed that the mean of the items corresponding to 
nursing students was slightly higher than the mean of the items about licensed nurses in each metric except for PI in which the items didn’t apply to nursing students.

\paragraph{Qualitative Feedback}

For the qualitative analysis, the scripts of the semi-structured interview were first transcribed from voice recordings to text. and then analyzed by the first author using thematic analysis method \cite{Braun2006}.
Specifically, the first author coded the sentences and paragraphs, then iteratively selected and grouped the codes. Then, the first and last authors reviewed all the codes to come up with the final themes. Examples of codes created at earlier steps of analysis included ``realistic patient audio feedback" suggests improvements on user experience and ``enable critical thinking," which suggests creating multiple difficulty levels. In the end, w\color{black}e identified five main themes in our preliminary findings:
\textbf{Theme 1: The medication system requires higher fidelity to real clinical workflows, including accurate timing and EHR documentation.}
Participants requested higher fidelity to real clinical systems, including accurate time for medications and integration of EHR for documentation practice. Some suggested that five minutes in VR should match the real time (P2), or let students navigate ordered medicines in a realistic EHR (P5). 
Post-simulation feedback reports were suggested to reinforce self-assessment, such as a summary of handwashing duration and choices made, etc. (P2).

\textbf{Theme 2: Simulations should reduce scripting to foster critical thinking through environmental assessment.}
Participants wanted more unguided decision-making with concerns that Students might just click all options—they need to assess the environment to decide, and complexity was appropriate but overly scripted (P1). Some recommended expanding scenario complexity with additional treatment steps and optional actions, noting this would better prepare students for clinical ambiguity by mirroring real-world decision-making challenges (P5). Some also suggested replacing button-based inputs with direct object manipulation such as clicking a pulse oximeter to apply it (P1) to better replicate clinical workflows in real life.

\textbf{Theme 3: UI design must balance competing preferences: consolidated data (for efficiency) vs. selective information (for focus).}
Participant preferences for physiological data displays varied: some advocated for consolidated monitoring approaches such as displaying all physiological data on a single monitor to minimize cognitive load, or prefer the room settings to be similar to their personal workspace in real life (P6),
while others endorsed selective information presentation in the current settings, arguing that simplified interfaces promote better concentration for students (P2).

\textbf{Theme 4: Participants Valued Verbal Patient Interactions but Praised Existing Audio Feedback for immersion.}
Participants emphasized the lack of verbal engagement opportunities, such as talking to the patients directly and asking for their feelings (P7), while they praised the audio from both the patient and the ambient surroundings, which made them feel realistic (P1). This contrast highlights both the demand for more naturalistic patient dialogue and the effectiveness of current auditory elements in sustaining engagement.

\textbf{Theme 5: VR training is perceived as accessible and effective for students' practice.}
Participants widely acknowledged the system's potential, praising its usability (P1-P8) and ability to enhance students' self-confidence while reducing anxiety (P5) because participants can make mistakes without being watched by their peers and instructors in person. They also highlighted its value as an accessible supplemental training tool that could complement traditional education methods (P8).

\section{Design Iteration of Framework: Phase 2}
\subsection{System Refinement and Expansion}
\subsubsection{Existing Scenario Modification}
\begin{figure*}
    \includegraphics[width=\columnwidth]{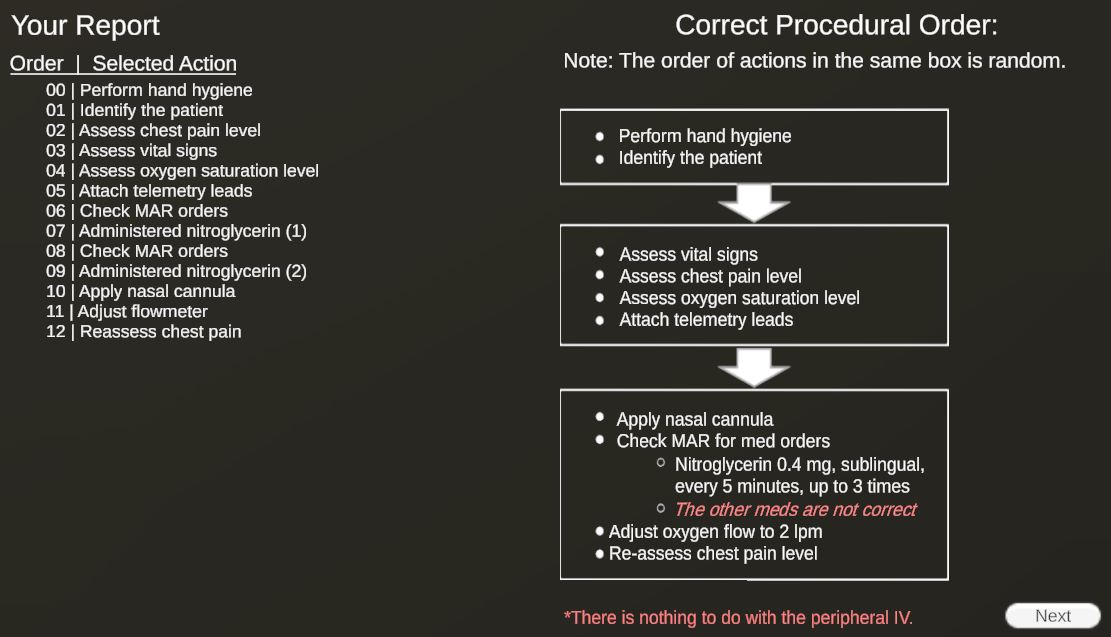}
    \hspace{.05\columnwidth}
    \includegraphics[width=\columnwidth]{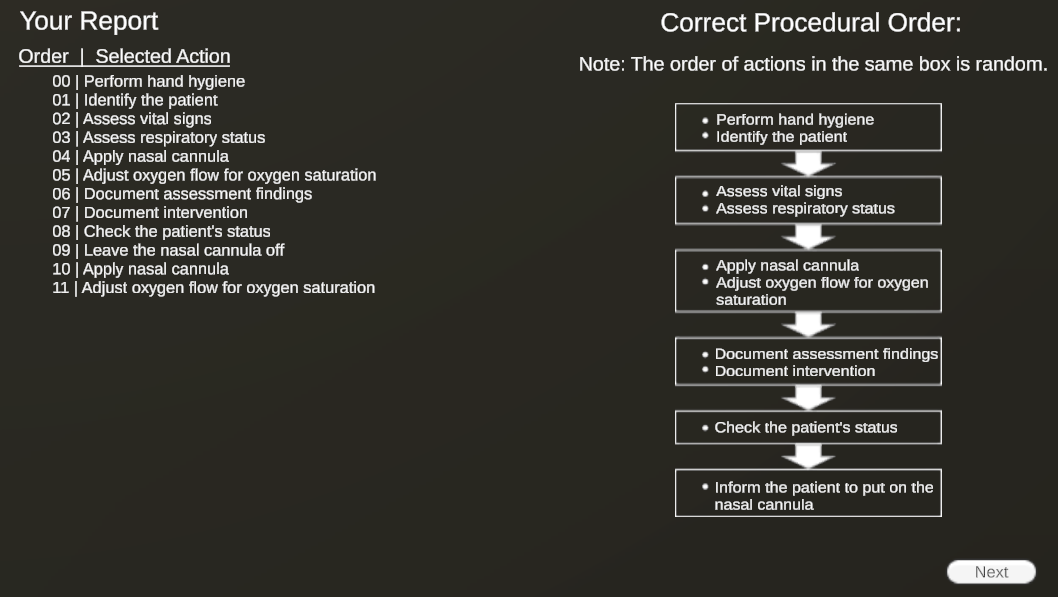}
    \caption{The layout of the final reports for the chest pain (up) and the oxygen therapy (down) scenarios in Study 2}
    \Description{Two example feedback reports list completed actions for the chest pain and oxygen therapy scenarios.}
\label{fig:report}
\end{figure*}

Following the identification of design considerations from nursing experts in Study 1, we collaborated with nursing specialists to systematically evaluate which suggested improvements from Study 1 to prioritize for implementation. 
Our decision-making process was guided by three key factors: (1) alignment with procedural skill training objectives, (2) the inherent difficulty level of the simulated scenarios, and (3) the development time required. 
This ensured that the final refinements balanced educational value with practical feasibility. 
The improvements included:
\begin{enumerate}
    \item Displaying the report of the order of actions the user selected at the end of the scenario description (see Fig. \ref{fig:report}). (Theme 1)
    \item Displaying all buttons on a panel attached to the non-dominant hand and providing a visual cue for where to perform an action, considering the potential cognitive load introduced by finding the buttons in an unfamiliar environment. (Theme 3)
    \item Adding a timing prompt to the chest pain scenario where time-sensitive tasks are involved. (Theme 1)
\end{enumerate}

\subsubsection{Expand The Current Scenario}
After we improved the chest pain scenario based on the feedback we received in Study 1, we designed and developed a second scenario with expanded complexity and more operations, as suggested in Theme 2.
Our nursing collaborators identified oxygen therapy as another frequently encountered intervention in acute care, motivating its selection as the new training scenario.
In the oxygen therapy scenario, the virtual patient has difficulty breathing and the goal of the simulation is to alleviate the symptoms. Similarly, the simulation concludes when the user completes all required treatments, indicated by the patient's verbal confirmation of symptom relief.

Transferrable features from the chest pain scenario include standard workflow such as hand hygiene and identifying patients' name, and interactions like operating the same devices and the patient's verbal feedback. 
New features include scenario-related treatments such as examining the patient's lung sounds, documentation of users' findings, and interventions through verbal report in an office room. 
Adjusting the oxygen flow speed is one of the key treatments in this oxygen therapy simulation. The patient would respond verbally to the speed as well as the oxygen saturation data as indicators for users to assess whether the symptoms were improving.
Whereas in the chest pain scenario where the key treatment is to administer medications, this was implemented using a progress bar for simplicity.
After coming back from the office room, users would find that the patient took off the nasal cannula, and they would need to decide whether to leave it off or ask the patient to put it back on.
To maintain similarity with real-life nursing scenarios, and to cultivate a mindset of critical precision and safety, users are not allowed to redo procedural steps once completed. 

Unity Coroutines provide essential flexibility for managing both automated and user-driven workflows of a dynamic interaction system.
The system dynamically adapts to task requirements: pre-built animations handle routine procedures to maintain workflow continuity, while manual intervention points suspend the state machine to accommodate user actions upon treatment selection.
This approach imposes no restrictions on the number or sequence of actions, ensuring dynamic and uninterrupted task execution.
We developed a new format to represent the linked lists of action sequences, with unique symbolic identifiers representing distinct animation and interaction types.
Adding new interaction types only requires implementing their symbolic handlers while maintaining existing workflow integrity (see supplemental materials for details).
Other new elements of the revised system include action layouts, manual interaction, time prompts, and reports (see Fig. \ref{fig:teaser}).

\subsection{Study 2: Effectiveness Study} \label{study2}
The purpose of this study was to further explore the utility and effectiveness of using this system in nursing education, mainly focusing on our target users - prelicensure nursing students. 
The order of the two scenarios was counterbalanced in this study. 

\subsubsection{Participants}
18 pre-licensure nursing students were recruited from the local School of Nursing for this study.
One was excluded due to technical issues occurred during the study.
These students were drawn from either a four-year bachelor's program or a two-year master's program and had varying levels of clinical training.
All had normal vision or vision corrected to normal by contact lenses.
One participant was male and 16 were female. 
Participants fell into the following age groups: 11 between 18-24, three between 25-34, and three between 35-44. 
Slightly less than half of participants (41.16\%) had at least one year of prior experience playing 3D video games. 

\subsubsection{Procedure \& Analysis}
The study procedure was the same as the first study except that users were told to familiarize themselves with the manual interactions in VR, and they answered open-end questions instead of being interviewed in the end. The total amount of time each participant was immersed in VR was about 30 minutes including the tutorial VR session. Each participant was compensated with a \$30 Amazon.com gift card.
The Creighton Competency Evaluation Instrument - Clinical Care Environments (CCEI-C) [anonymized citation] was used to evaluate students' competency and performance, which consists of a check list of treatments a student has to perform in the corresponding scenario.
The recorded treatment order of each participant was analyzed by a nursing educator specialist (the second author). 
The higher the score, the better the students performed.
The Post-Study System Usability Questionnaire (PSSUQ) and the After-Scenario Questionnaire (ASQ) \cite{PSSUQ} were used to evaluate the usability and users' satisfaction of this simulation. Three items in the original PSSUQ were removed because they were either duplicated with previous questions or not applicable in our simulation. 
The item \textit{``I believe I could become productive quickly using this system"} in PSSUQ was redundant with \textit{``Using the simulation app would increase my productivity"} in the UAQ.
\textit{``The system gave error messages that clearly told me how to fix problems"} was excluded because our simulation intentionally omitted error messages to avoid interrupting training scenarios.
\textit{``Whenever I made a mistake using the system, I could recover easily and quickly"} was removed to better reflect real clinical practice where mistakes have cumulative consequences without reversal options.

Similar to Study 1, the SSQ and UAQ were used to assess cybersickness and user acceptance, respectively.
The PEOU subscale in the UAQ was removed from the UAQ due to overlaps with items in the PSSUQ like \textit{``It was simple to use this system"}, \textit{``It was easy to learn to use this system"}, \textit{``It was easy to find the medical devices and information"}. The modified questionnaires used in the study are included as supplemental material.

After each scenario, a question ``How confident do you feel about the order of the operations you took in this scenario?" was asked before the report of users' action order was displayed, and ``How confident do you feel you will choose the correct order of operations in this scenario when you do it \textit{next time}?" was asked after the report. 
In addition, four relevant items about content mastering, necessary content coverage, skills and knowledge and usage for learning in the National League for Nursing (NLN)'s questionnaire were selected and used in the post-study survey \cite{NLN}. 
At the end of the post-study survey, there were three open questions asking what participants liked and disliked about the two simulations and any open feedback they had. 
Demographics information was collected at the end of the session.
Participants were not asked to repeat the simulation after reviewing their performance due to the time constraints of each study session. 

\subsection{Study 2: Results}
\subsubsection{Quantitative Results}
\paragraph{Performance}
The mean score of the oxygen therapy scenario was 5.94 out of 7 ($\approx85\%, Mdn=6, SD=0.9$), and 7 out of 8 ($\approx87\%, Mdn=7, SD=1.22$) for the chest pain scenario.
The mean time spent on the chest pain simulation was 343.26 seconds ($SD$ = 118.21s), and 547.59 seconds ($SD$ = 136.88s) on the oxygen therapy scenario. 
According to the feedback provided by our nursing education specialists, scores of 85–87\% are considered good outcomes for these types of training simulations.
The time duration was longer than the formative study because the modified version had more manual interactions. 
The time cost of the oxygen therapy was also longer than the chest pain scenario because it had more interactions and a longer conversation.
Even though the default order of the actions on the panel was the correct order, not all participants performed them correctly, which indicates that the difficulty level of the simulation was still challenging. 
For example, the action ``perform peripheral IV'' was a distraction in the chest pain scenario, while some participants still chose that option when the patient reported severe chest pain. 
In the oxygen therapy scenario, the nasal cannula needed to be left on the patient after the oxygen therapy, but some participants removed it. 
We also observed that some participants simply selected actions in the default order presented to them.

\begin{figure}[htbp]
     \centering
     \includegraphics[width=0.7\linewidth]{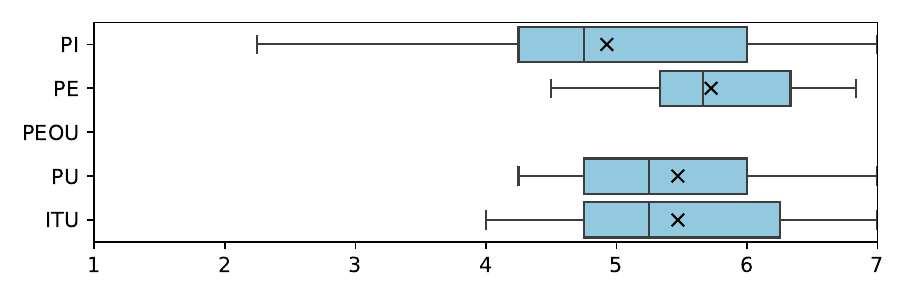}
     \caption{Box plots of UAQ-subscale measures from nursing students in Study 2, displaying the mean (cross marks), median, and IQR. }
     \Description{Horizontal box plots summarize five user acceptance subscales for nursing students in Study 2.}
\label{fig:uaq-subscale}
\end{figure}

\begin{figure}[htbp]
     \centering     
     \includegraphics[width=0.7\linewidth]{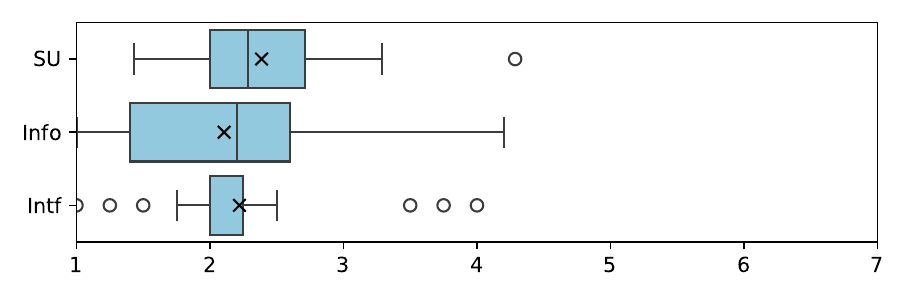}
     \caption{Box plots of PSSUQ-subscale measures from nursing students in Study 2: System Usefulness (SU), Information Quality (Info), Interface Quality (Intf), displaying the mean (cross marks), median, and IQR.}
     \Description{Horizontal box plots summarize system usefulness, information quality, and interface quality ratings in Study 2.}
\label{fig:pssuq-subscale}
 \end{figure}
 
\paragraph{User Acceptance \& Usefulness}
The mean of the PI, PE, PU, and ITU in the UAQ was 4.93 ($SD$ = 1.33), 5.73 ($SD$ = 0.68), 5.47 ($SD$ = 0.88), 5.47 ($SD$ = 0.83) respectively (see Fig. \ref{fig:uaq-subscale}).
The PE, PU, ITU subscales in the UAQ showed high perceived enjoyment, usefulness and intention to use. 
The PI subscale was low, which was consistent with participants' video game experience. 
The overall PSSUQ score was 2.26 ($SD$ = 0.71). 
The score of the system usefulness, information quality, interface quality subscales was 2.39 ($SD$ = 0.75), 2.10 ($SD$ = 0.79), and 2.22 ($SD$ = 0.82) respectively (see Fig. \ref{fig:pssuq-subscale}). The questions in the PSSUQ were customized and slightly different from the original version, but the results still showed high usability and utility of this simulation \cite{SAURO2016185}.

\paragraph{Satisfaction}
The ASQ score of the chest pain scenario was 2.37 ($SD$ = 0.77) and 2.08 ($SD$ = 0.61) of the oxygen therapy scenario.   
Although we did not find any existing benchmarks of the ASQ score from previous literature, the result still to some extent showed that participants were highly satisfied with the ease of completing this task, the amount of time it took to complete this task, and the support information (such as the tutorial, instruction panels, visual hints) when completing this task, even there were some confusions in the instructions as reported in the open questions. 
This is consistent with the perceived enjoyment and engagement from the qualitative feedback.

\begin{figure}[htbp]
     \centering     
     \includegraphics[width=0.7\linewidth]{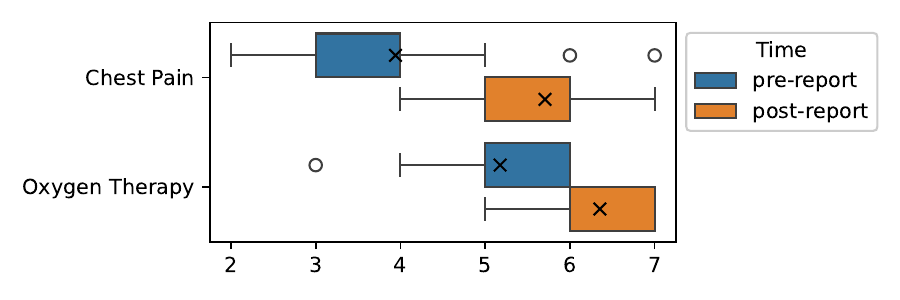}
     \caption{Box plots of self-confidence scores after the chest pain and oxygen therapy scenarios in Study 2, displaying the mean (cross marks), median, and IQR. In both scenarios, the confidence scores were significantly higher after viewing the report compared to before.}
     \Description{Horizontal box plots compare nursing students self-confidence before and after viewing reports for both scenarios.}
\label{fig:as-confidence}
 \end{figure}

\paragraph{Self-Confidence}
In after-scenario surveys, the range of the scale of the self-confidence items is from 1 (not at all) to 7 (very confident). 
The mean of pre-report and post-report was 3.94 ($Mdn$ = 4, $SD$ = 1.25) and 5.71 ($Mdn$ = 6, $SD$ = 1.05) in the chest pain scenario, and 5.18 ($Mdn$ = 5, $SD$ = 0.81) and 6.35 ($Mdn$ = 7, $SD$ = 0.86) in the oxygen therapy scenario (see Fig. \ref{fig:as-confidence}). 
The Wilcoxon Signed-Rank Test was used due to the normality violation indicated by the Shapiro-Wilk normality test. 
A significant difference was found between pre-report confidence and post-report confidence in both the chest pain ($z$ = -3.45, $p <$ 0.001) and the oxygen therapy scenario ($z$ = -3.22, $p <$ 0.001). 
In the post-survey, a 7-point Likert scale was used for the self-confidence questions instead of a 5-point scale to keep consistent with the other questions. The scale ranges from 1 (strongly disagree) to 7 (strongly agree). The overall mean score was 5.90 ($SD$ = 0.59).
The after-scenario self-confidence ratings significantly increased in both scenarios, which indicates that having access to the reports and the correct answers can help participants build their confidence and enhance self-assessment skills. 
The ratings in the post-study survey also indicates a overall high self-confidence level. 

\paragraph{Cybersickness}
Only one participant self-identified motion sick, both before and after the study. 
The mean of the delta overall SSQ scores is 8.80 ($Mdn$ = 3.74, $SD$ = 11.67). 
The Wilcoxon Signed-Rank Test was used due to the normality violation as indicated by the Shapiro-Wilk normality test. 
A significant difference was found between the pre-study and the post-study SSQ scores ($z$ = -2.82, $p$ = .002).
The SSQ results indicated a significant, albeit mild, increase in symptoms after the VR experience \cite{SSQInterpretation}, which is not surprising given the total amount of the time they were immersed in VR.  
Participants reported eye strain and difficulty focusing, which was consistent with the feedback that they had blurry vision.

\subsubsection{Participant Feedback}
\paragraph{Practice and Training Benefits}
Most participants thought this nursing simulation was a helpful practice, such as verbal documentation, directed tutorial and learning and feedback on the actions. 
They also felt it was useful to have a list of actions to choose from, which made it simple to understand and allowed them to think through different scenarios. 
One participant pointed out that they could make choices independently whereas they always had a partner to do a teamwork, and also it was okay to make mistakes in low stakes environments like this virtual simulation. 

\paragraph{Strong Engagement in Immersive Interfaces}
Some participants also enjoyed the realistic settings, instruments and patient scenarios. 
One participant mentioned that the simulation was helpful in the way that the distracting components in the simulation such as the ambient sound and navigation were also challenging in real-life situations, and students could get used to it by practicing in this simulation. 
Another participant thought it was more realistic than a mannequin in terms of the patient's feedback and healthcare setup, and less realistic due to not being in a physical space and less tactile working with controllers. 
Some participants also expressed their appreciation of the way to interact with patients, such as putting on the BP cuff, checking the patient's lung sounds, adjusting oxygen flow meter, communicating with the patient, etc.

\paragraph{VR Acclimatization Challenges \& Comfort Limitations}
It took a while for some participants to get familiar with the simulation due to various reasons. 
For example, two participants mentioned that they had an unclear eyesight at the beginning even though they adjusted the headset and the IPD beforehand and confirmed that they were ready to start. 
Some participants reported discomfort after wearing the headset for a long time, including blurry vision and heavy weight, which remains a limitation for many commercially available VR headsets.

\paragraph{Usability Friction Points}
There were still some confusions participants identified in the simulation, such as being not sure whether to verbally talk to the patient or not, not knowing there were unnecessary actions, having difficulty locating the green arrow, inconsistency of the operations of adjusting the oxygen flow meter. 
However, these can mostly be attributed to the overhead of learning a new interface and lack of familiarity with VR systems. Two participants reported that they initially did not know the next step but were able to figure it out later by themselves.
We believe that the overhead of learning new interfaces and simulation will be largely reduced when students get familiar with the VR system and when this simulation is introduced by instructors along with a course and used as supplemental materials.
In terms of locomotion, one participant preferred continuous rotation over snap turning. 
One also had difficulty locomoting, while another found it easy to grab objects and move around. 
One participant felt uncomfortable to put on the headset due to its heaviness, and another felt dizzy after the simulation. 
In the future, physical discomfort from headset weight and ergonomics could potentially be mitigated by using an after-market head strap.

\paragraph{Potential Improvements}
Some participants preferred a real Medication Administration Record (MAR) or Electronic Health Records (EHR) system. 
In the current simulation, they could not access the detailed medicine history of the patient when they gave the medicine order. 
Some also wanted to have more dialogues and interactions with the patient, as well as a more interactive way of adjusting the oxygen flow meter. 
One participant hoped to access the rationale of the correct orders. 
Another participant preferred clicking through the buttons and felt tedious of physical tasks. 
One also noted that it was helpful to have someone guide through the simulation if they encountered any operation problems.
Some participants argued that this simulation could be utilized for teaching nurse classes, as there is no need of time of resources like the standard patient simulation and no risk of dealing with an actual patient. 
It could help students feel better prepared before going to the real clinic and be a good refresher for practice.

\section{Discussion} \label{sec:discussion}
\subsection{Research Implications}
\paragraph{Generality of VR training systems}
Many VR nursing training studies remain narrowly focused on scenario-specific implementations. 
In contrast, our work demonstrates how iterative design, technical implementation, and explicit design choices can support the development of modular nursing VR systems that extend across multiple procedural contexts. For the research community, it is essential to capture and share considerations from all stakeholders—educators, clinicians, and technologists alike.
We therefore emphasize the importance of transparent reporting of both pedagogical and technical design decisions, as this enables replication, facilitates cross-institutional validation, and advances the generalizability of VR training research.

\paragraph{Influence of Iterative Design on Clinical Skill Development}
In this study, we evaluated user acceptance, satisfaction, and self-confidence, which are metrics commonly emphasized in prior nursing VR research. 
Beyond these outcomes, our work demonstrates that iterative co-design with domain experts is not simply a development strategy, but a critical factor in shaping simulation fidelity, task sequencing, and the alignment of training with real-world clinical practice. 
Iterative engagement allowed us to refine interactions, integrate authentic workflows, and balance educational goals with system usability. 
This highlights the broader research implication that VR training systems benefit most when design is treated as an evolving dialogue between educators, domain experts, and technologists. 
Future studies should therefore move beyond self-reported learner outcomes to systematically investigate how specific design choices (e.g., guided versus unguided decision-making, level of scenario scripting) influence not only clinical reasoning and error management, but also learners’ ability to transfer skills into authentic practice.

\paragraph{Temporal interaction element}
Simulating the passage of time in VR is another noticeable challenge, yet still underexplored in existing research work. 
In the chest pain scenario, the nitroglycerin was administered every 5 minutes, and a text prompt was visualized on the monitor after a few seconds. 
However, we observed that some participants did not notice this prompt.  
Additionally, the time becomes more difficult to convey when multiple medications are being administered over different intervals.
Existing approaches include a visualized clock in the virtual environment, some incorporate ticking sound. 
However, for cases where the patient needs to be treated beyond the time of a study section, it is still difficult to track and evaluate students' performance if one needs a realistic simulation. 
The question of how to implement and design the elapsed time remains unexplored when implementing the time-sensitive interactions or treatments that need to be performed periodically in the research community. 
In the future, a more conspicuous method that combines visual and audio feedback would be helpful to ensure that participants are aware of the passage of time, especially when it involves administering medications according to a schedule.

\paragraph{Interdisciplinary collaboration}
In the developmental phase, it was observed that nursing collaborators possessed a deep understanding of users' behaviors and patients' responses. 
However, they exhibited limited proficiency in accounting for all potential scenarios within the dialogue tree, a strength typically associated with technical professionals. 
Conversely, technical experts excelled in this aspect, but demonstrated comparatively less insight into users' actions and patients' reactions. 
It remains a challenge of designing the dialogue system with domain experts. We need to go through all details of the logic and testing. 
Future integration of large language models with medical reasoning systems could reduce development time and optimize dialogue complexity, as demonstrated by real-time communication experiments \cite{10536583}. 

Through the collaboration in the two phases and interviews from the expert evaluation study, we noticed that nursing experts may have limited knowledge about what the VR technology can achieve. 
Sometimes we need to give nursing experts an introduction to the technology to provide more context for them to offer suggestions.
Also, a student preferred continuous turning over snap turning, however, snap-turning is commonly used in commercial VR games to mitigate cybersickness \cite{PPR:snap-turning}, but participants that have not used it before may prefer continuous rotation.
We argue that when collaborating with nursing experts who unfamiliar with the technology, providing some brief explanation and hands-on testing would be beneficial for more accurate and concise design considerations.
For the research community, this suggests that VR in healthcare education should be studied not only as a training tool but also as a design process shaped by the dialogue between diverse stakeholders.

\subsection{Design Considerations}
\textbf{\textit{Balance guidance and autonomy.}}
Some difference in the considerations of design priorities occurred when comparing the feedback from experts and students.
We found that participants prefer to practice in familiar clinical environmental settings especially when they have previous experience with some real-life clinical sites. 
Although the goal of the training system mainly focuses on the medical knowledge aspect, experienced nurses prefer doing experiments in the environment identical to their local sites. 
Because the specific environment used may influence learning outcomes \cite{flott2016clinical}, designers may consider replacing the 3D assets in the training system to fit the clinical environment depending on the target users.

\textbf{\textit{Accommodate diverse learner needs.}}
We also observed contrasting preferences for instructional support. Although participants in both studies enjoyed the manual interaction, some nursing experts prefer removing all text prompts in the scenario and letting students operate directly with the devices to enhance their critical thinking and decision-making ability, while most students prefer the step-by-step break down of the scenario, which greatly facilitates getting started of the simulation from the beginning. 
In our system, we used an online asset and customized the medical devices that are necessary in the scenario, because our target user is the pre-licensure nursing students. 
For broader use cases, We recommend considering the target users' professional level, clinical experience and habits, and offering adaptive difficulty modes that gradually transition from structured support to autonomous practice.

\textbf{\textit{Support authentic clinical workflows.}}
Participants emphasized that the inclusion of realistic workflows enhanced their perception of realism and relevance. 
One nursing expert thought the VR experience was similar to the online computer-based simulation from the perspective of content design, and suggested that students might become less interested after using it for a while. 
However, it is worth pointing out that some manual interactions, like checking the lung sounds of the patient, provide opportunities for active engagement that can potentially enhance the performance of the learning.  
Additionally, some casual interactions, such as performing hand hygiene, require spending a certain amount of time to do it properly, which was highly valued by some nursing experts as they helped students connect the simulation with real-world practice. We therefore recommend considering embedding authentic workflow elements to strengthen skill transfer.

\textbf{\textit{Address VR-specific usability challenges.}}
The findings from both studies suggest that minimizing adverse experiences may require periodic breaks and proper headset adjustments. 
We suggest limiting immersion time to no longer than a half hour \cite{VR-duration} and providing opportunities to pause and take breaks. 
In addition, we observed that some participants frequently scanned the virtual environment for visual cues, which appeared to contribute to dizziness. 
While such behaviors may diminish with repeated exposure and increased familiarity, we noticed in the second study that participants using swivel chairs often rotated physically rather than relying on joystick controls, 
whereas participants were seated in a static chair in the first study.
However, since this system is designed for unsupervised use in personal spaces, we cannot standardize or control users' physical setups to isolate this effect. 
These observations collectively highlight the need for clear ergonomic guidelines while acknowledging the inherent limitations of self-directed VR training environments.
Designers should implement best practices from VR HCI research, including clear ergonomic guidelines, visual comfort settings, and minimal interface clutter.

\subsection{Limitations}
The studies presented in this paper have several limitations. 
First, the amount of participants was small, as recruiting participants from specialized groups is more challenging than the general population. 
Participants were mostly female; however, this gender distribution aligns with the demographics of the nursing field.
Additionally, while our nursing collaborators indicated that the training performance was satisfactory, baseline performance data was not collected before the training simulation. 
Although this simulation was designed to serve as a supplemental training tool for nursing education, it was not possible to deploy the simulation within an active classroom for logistical reasons.
Furthermore, a longitudinal study is needed in the future to formally evaluate the learning outcomes of these training scenarios when used over a longer period of time.

The initial version of the system only had progress bars instead of manual interactions. 
Although it was because the key element in the first scenario was to assess the patient's chest pain level from their verbal report, and administer the medicines accordingly, the progress bars make the interaction more tedious when designing the second scenario which requires more operations. 
The final version of the system also has several potential flaws that can be improved, such as inconsistency in similar features with different interactions in each scenario. For example, in the oxygen therapy scenario, one of the essential challenges involves adjusting the oxygen flow meter speed correctly to help the patient breathe.  
However, in the chest pain scenario, the main focus was the medicine order, so the manual speed adjustment was not enabled. 
This inconsistency in operating the virtual equipment made some participants feel confused. 
In the future, pop-up hints could be displayed if the user attempts to interact with an object in an appropriate way.

In these scenarios, the default action order presented to the user was actually the correct order.  Most students found that having pre-defined buttons was very helpful, and some students still made mistakes. 
However, some instructors thought the action buttons could be removed so that students could interact with the environments directly, which would substantially increase the difficulty. 
As indicated by some nursing experts, the simulation can be designed with multi-level difficulties for students at different professional levels.
Introducing more opportunities for randomization could increase difficulty and encourage repeat use.

\section{Conclusion and Future Work}
In this research, we developed a VR nursing system iteratively with a chest pain scenario and extended to oxygen therapy scenario which involves more interaction, aiming for nursing students to practice procedural skills. 
These scenarios require learners to recognize symptoms, follow protocols, and determine appropriate interventions through applied medical knowledge.
We conducted two studies to explore the feasibility and effectiveness, and the feedback from both nursing experts and students indicates that this simulation has a positive acceptability, usability, and utility in nursing education. 
We also provided guidelines for design of procedural skill training simulations for researchers in the related areas. Future work may consider evaluating students' learning outcomes with more complex scenarios experienced over longer periods of time,
comparing immersive and conventional training modalities, evaluating learning outcomes compared to baseline performance and/or after multiple rounds of training,
\color{black}
and investigating the transfer of knowledge to real-life clinical experiences.
\bibliographystyle{ACM-Reference-Format}
\bibliography{main}

\end{document}